\documentclass[12pt]{article}
\newcommand{\qqbar}{\bar q^{\!\!\!\!\!\textsuperscript{\tiny{(~\,\,)}}}}
\newcommand{\ppbar}{\bar p^{\!\!\!\!\!\textsuperscript{\tiny{(~\,\,)}}}}
\usepackage{epsfig}
\title{\bf QCD corrections to forward-backward charge asymmetries in
$l^-l^+j$ 
production at hadron colliders
\footnote{Talk presentad by R. Pittau at the {\em 
International Europhysics Conference on High Energy Physics},
		 July 21st - 27th 2005
		 Lisboa, Portugal}}
\author{
{\bf R. Pittau} 
\thanks{Work supported in part by MCYT under 
contract FPA2003-09298-C02-01 and by 
Junta de Andaluc{\'\i}a group FQM 101 and by
MIUR under contract 2004021808\_009.} \\
U. of Granada, CAFPE, U. of Torino and INFN Torino, Italy\\
E-mail: roberto.pittau@to.infn.it\\\\
{\bf F. del Aguila}\\
U. of Granada and CAFPE, Spain\\\\
{\bf Ll. Ametller}\\
 U. Polit\`ecnica de Catalunya,
Barcelona, Spain 
}

\date{}
\begin{document}
\maketitle
\begin{center}
{\bf \large Abstract}
\end{center}
\small
The large cross sections for gauge boson production at the 
Fermilab Tevatron and the CERN Large Hadron Collider (LHC) 
might give a chance to determine the electroweak parameters 
with high precision.
We calculated two different forward-backward charge 
asymmetries (${\rm A}^{\rm CS}_{\rm FB}$ and ${\rm A}^j_{\rm FB}$) 
of lepton pairs in events with a 
large transverse momentum jet 
$p \ppbar \rightarrow Z, \gamma ^* + j \rightarrow e^-e^+ + j$
at next-to-leading order (NLO), 
$\cal O(\alpha _{\rm s})$ corrections, making use of 
the Monte Carlo programs MCFM \cite{MCFM} and ALPGEN \cite{ALPGEN}.
These observables could provide a new determination of the weak 
mixing angle $\sin ^2 \theta _{\rm eff}^{\rm lept}(M_Z^2)$ with a statistical 
precision for each lepton flavour of $\sim  10^{-3}\ (7\times 10^{-3})$ 
at LHC (Tevatron). 
If $b$ jets are $\mbox{identified}$, a new asymmetry with respect 
to the {\it b} quark (${\rm A}_{\rm FB}^b$)
can also be measured with a 
statistical precision of $\sim  2\times 10^{-3}\ (4\times 10^{-2})$ 
at LHC (Tevatron).
Finally, we comment on the dependence of our results on various sources
of uncertainties and compare, in the case of ${\rm A}_{\rm FB}^b$,
the exact result with an approximation that might be
more suitable when performing a realistic experimental analysis.
\clearpage

\normalsize
\section{The definition of the forward-backward charge asymmetries}
As explained in \cite{delAguila:2005cn},
the optimal observable to quantify 
possible correlations in the direction of emission
of the {final} state lepton in the process
$
p \ppbar \rightarrow Z, \gamma ^* + j \rightarrow e^-e^+ + j
$,
is a forward-backward asymmetry:
\begin{eqnarray}
{\rm A}_{\rm FB} = \frac {F-B}{F+B}\,~~~~\mbox{with}~~ 
F = \int _0^1 \frac{{\rm d}\sigma}
{{\rm d}\cos \theta }
{\rm d}\cos \theta\, ~~\mbox{and}~~ 
B = \int _{-1}^0 \frac{{\rm d}\sigma}
{{\rm d}\cos \theta }
{\rm d}\cos \theta \,. \nonumber
\end{eqnarray}
One can consider two possible angles:
\begin{eqnarray}
\cos \theta _{\rm CS} &=&
\frac{2 (p_z^{e^-}E^{e^+}-p_z^{e^+}E^{e^-})}
{\sqrt {(p^{e^-} + p^{e^+})^2}
\sqrt {(p^{e^-} + p^{e^+})^2 + (p^{e^-}_T+p^{e^+}_T)^2 }}\,~~\mbox{or}~~ 
\nonumber \\
\cos \theta _j &=& 
\frac{(p^{e^-}-p^{e^+})\cdot p^j}
{(p^{e^-}+p^{e^+})\cdot p^j}\,, \nonumber 
\label{costheta} 
\end{eqnarray}
where the four-momenta are measured in the 
laboratory frame and $p^{\mu}_T \equiv (0,p_x,p_y,0)$.
The Collins-Soper angle $\theta _{\rm CS}$ is, 
on average, the angle between $e^-$ and the initial quark direction,
while $\theta _j$ is the angle between 
{$e^-$} and the direction opposite to the jet in the 
{$e^-e^+$} rest frame \cite{Aguila02}.

Different asymmetries can then be defined, according to 
the scheme given in Table \ref{table:1a}.
\begin{table}[ht]
\begin{center}
\begin{tabular}{|c||c|l|}
\hline
Collider & Asymmetry & ~~~~~~~~~~~Definition \\ \hline \hline
$p \bar p $ & ${ A^{\rm CS}_{\rm FB}}$ & 
$\cos \theta = \cos \theta_{\rm CS}$    \\
$p      p $ & ${ A^{\rm CS}_{\rm FB}}$& $\cos \theta = \cos \theta_{\rm CS} 
 \times \frac{|p_z^{e^+}+p_z^{e^-}+p_z^j|}{p_z^{e^+}+p_z^{e^-}+p_z^j}$ 
       \\ \hline
$p \bar p $ & ${ A^{j}_{\rm FB}}$ & $\cos \theta = \cos \theta _j 
     \times \frac{|p_z^{e^+}+p_z^{e^-}+p_z^j|}{p_z^{e^+}+p_z^{e^-}+p_z^j}$ \\ 
$p      p $ & ${ A^{j}_{\rm FB}}$ & $\cos \theta = \cos \theta _j$ \\ \hline
$p \bar p $ & ${ A^{b}_{\rm FB}}$ & $\cos \theta = \cos \theta _j 
             \times (-{\rm sign}(Q_b))$ \\
$p      p $ & ${A^{b}_{\rm FB}}$ & $\cos \theta = 
\cos \theta _j\times (-{\rm sign}(Q_b))$ 
\\ \hline
\end{tabular}
\end{center}
\caption{\label{table:1a} \small
The definitions of the various asymmetries at the $pp$ and $p \bar p$
colliders.}
\end{table}
The extra phase factors in the definitions of Table \ref{table:1a}
ensure non vanishing asymmetries. Notice that, in the case 
of ${\rm A}^{b}_{\rm FB}$, one must detect the charge 
of the produced $b$ jet. 
\section{Numerical results and conclusions}
\begin{figure}[h]
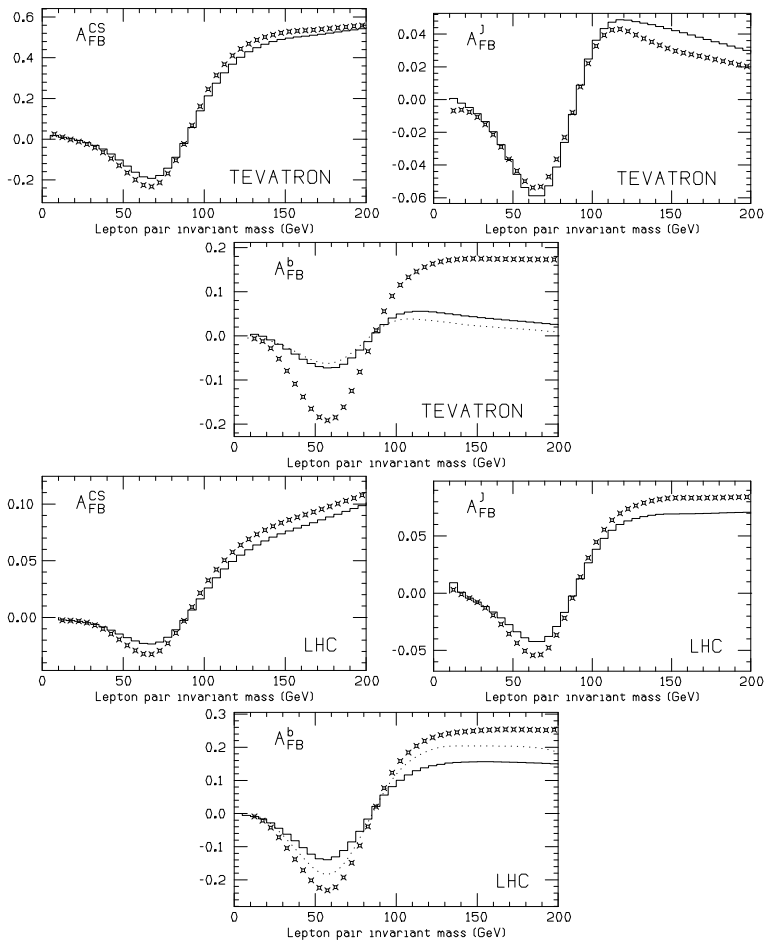

\begin{center}
\epsfig{file=Acs_tev_conl.eps,height=4.97cm,angle=90} 
\epsfig{file=Aj_tev_conl.eps,height=4.97cm,angle=90} \\
\epsfig{file=Ab_tev_proc.eps,height=4.97cm,angle=90} \\
\epsfig{file=Acs_lhc_conl.eps,height=4.97cm,angle=90}
\epsfig{file=Aj_lhc_conl.eps,height=4.97cm,angle=90} \\
\epsfig{file=Ab_lhc_proc.eps,height=4.97cm,angle=90}
\end{center}
\caption{\label{fig:1} \small
NLO (solid histogram) and LO (points) 
asymmetries at Tevatron and LHC.
The dotted lines refer to the approximation of using 
only ALPGEN, as described in the text, with $m_b= 4.62$ GeV.}
\end{figure}

The effect of including NLO $\cal O(\alpha _{\rm s})$ corrections
in the production rates is given in Table \ref{table:2}, while
the corresponding change in the asymmetries is depicted in Figure \ref{fig:1}.
\begin{table*}[ht]
\begin{center}
\begin{tabular}{|c||c|c|c|c|}
\hline
{\rm Contributing} &  \multicolumn{2}{c|}{\rm LHC}
  & \multicolumn{2}{c|}{\rm Tevatron} \\
{\rm process} & \multicolumn{1}{c}{\rm LO} & \multicolumn{1}{c|}{\rm NLO}
  & \multicolumn{1}{c}{\rm LO} & \multicolumn{1}{c|}{\rm NLO}\\
\hline
\hline
$g \qqbar \rightarrow Vj(j)$ & 44.3 & 53.4 
  & 3.40 & 4.77 \\
$\begin{array}{c}
q\bar q \rightarrow Vj(j) \\
\qqbar \qqbar \rightarrow Vj(j) \\
g g \rightarrow Vj(j) \\
\end{array}$ & 
$\begin{array}{c}
8.4 \\
- \\
- \\
\end{array}$ & 
$\left.
\begin{array}{c}
 \\
 \\
 \\
\end{array}
\right\} 3.7$ 
& $\begin{array}{c}
  4.61 \\
  - \\
  - \\
  \end{array}$ & 
  $\left.
  \begin{array}{c}
   \\
   \\
   \\
  \end{array}
  \right\} 2.76$ \\
\hline
{\rm Total} & 52.7 & 57.1 
  & 8.01 & 7.53 \\
\hline
\hline
$\begin{array}{c}
gb \rightarrow Vb(g) \\
gg \rightarrow Vb(\bar b) \\
\qqbar b \rightarrow Vb(\,\,\qqbar) \\
\end{array}$ &
$\begin{array}{c}
1.81 \\
- \\
- \\
\end{array}$ &
$\left.
\begin{array}{c}
 \\
 \\
 \\
\end{array}
\right\} 1.81 $ 
  & $\begin{array}{c}
  0.038 \\
  - \\
  - \\
  \end{array}$ &
  $\left.
  \begin{array}{c}
   \\
   \\
   \\
  \end{array}
  \right\} 0.049 $ \\
$q\bar q \rightarrow Vb(\bar b)$ & $-$ & 0.06 
  & $-$ & 0.025 \\
\hline
{\rm Total} & 1.81 & 1.87 
  & 0.038 & 0.074 \\
\hline
\end{tabular}
\end{center}
\caption{\label{table:2} \small Estimates for the $e^-e^+j$ and $e^-e^+ b$ 
cross sections at LHC ($\sqrt s = 14$ TeV)
and Tevatron ($\sqrt s = 1.96$ TeV) in pb. 
The jet transverse momenta are required to be larger than 
50 (30) GeV at LHC (Tevatron)
and all pseudorapidities $|\eta |$ smaller than 2.5. 
The $p_t$ of the leptons is larger than 20 GeV.
The separations in the pseudorapidity-azimuthal angle 
plane satisfy $\Delta R > 0.4$ 
and $M_{e^-e^+}$ is within the range $[75,105]\, {\rm GeV}$. 
$\qqbar$ means summing over $q$ and $\bar q$  contributions
and $V \equiv Z, \gamma ^*$.}
\end{table*}     
%
A measurement of such asymmetries can be directly translated into a 
determination of $\sin ^2 \theta _{\rm eff}^{\rm lept}(M_Z^2)$.
The precision reach $\delta \sin ^2 \theta _{\rm eff}^{\rm lept}$ 
is given in  Table \ref{table:3},
assuming an integrated Luminosity  {$L$} of
100 {(10)} $fb^{-1}$ at LHC {(Tevatron)}.
\begin{table*}[ht]
\begin{center}
\begin{tabular}{|c|r|c|c|c|c|c|}
\hline
\begin{tabular}{c}
LO 
 \\ NLO
\end{tabular}
& $\sigma({\rm pb})~~~~$ &   &  ${\rm A}_{\rm FB}$ & 
$\delta{\rm A}_{\rm FB}$ & $\delta \sin ^2 \theta _{\rm eff}^{\rm lept}$ \\
\hline
\hline
LHC      & $\sigma^{Vj}=$ 53  & $A^{\rm CS}_{\rm FB}$ & $8.7 \times 10^{-3}$ & $4.4 \times 10^{-4}$ & $1.3 \times 10^{-3}$ \\ 
         &      {  57} &          & ${  6.8 \times 10^{-3}}$ & ${  4.2 \times 10^{-4}}$ & ${  1.3 \times 10^{-3}}$ \\
         &                    & $A^{j}_{\rm FB}$  & $1.2 \times 10^{-2}$ & $4.4 \times 10^{-4}$ & $8.8 \times 10^{-4}$ \\
         &               &          & ${  1.1 \times 10^{-2}}$ & ${  4.2 \times 10^{-4}}$ & ${  1.1 \times 10^{-3}}$ \\
         & $\sigma^{Vb}=$ 1.8 & $A^{b}_{\rm FB}$  & $7.5 \times 10^{-2}$ & $2.3 \times 10^{-3}$ & $8.7 \times 10^{-4}$ \\ 
         &          {  1.9}&          & ${  4.9 \times 10^{-2}}$ & ${  2.3 \times 10^{-3}}$ & ${  1.4 \times 10^{-3}}$ \\ 
       \hline \hline
Tevatron & $\sigma^{Vj}=$ 8.0 & $A^{\rm CS}_{\rm FB}$ & $6.4 \times 10^{-2}$ & $3.5 \times 10^{-3}$        & $1.4 \times 10^{-3}$ \\ 
         &          {  7.5}&   & ${  5.5 \times 10^{-2}}$ & ${  3.6 \times 10^{-3}}$ & ${  1.7 \times 10^{-3}}$ \\
         &                    & $A^{j}_{\rm FB}$  & $      9.9 \times 10^{-3}$ & $3.5 \times 10^{-3}$ &        $8.1 \times 10^{-3}$ \\
         &                    &   & ${  1.1 \times 10^{-2}}$ & ${  3.6 \times 10^{-3}}$ & ${  7.2 \times 10^{-3}}$ \\
         & $\sigma^{Vb}=$ 0.04& $A^{b}_{\rm FB}$  & $      5.5 \times 10^{-2}$ & $5.1 \times 10^{-2}$ &        $2.5 \times 10^{-2}$ \\ 
         &         {  0.07}&   & ${  2.7 \times 10^{-2}}$ & ${  3.7 \times 10^{-2}}$ & ${  4.7 \times 10^{-2}}$ \\
\hline
\end{tabular}
\end{center}
\caption{\label{table:3} \small Estimates 
for the $e^-e^+j$ and $e^-e^+ b$ 
cross sections and asymmetries defined in the text
with $M_{e^-e^+}$ in the range $[75,105]\, {\rm GeV}$.
The first row of each entry is the LO result, while the second one 
is the NLO. 
The integrated luminosity and the cuts can be found
in the text.
The statistical precisions are also given.}
\end{table*}     
The size of the NLO corrections is moderate, except  
for $\sigma^{Vb}$ at Tevatron in Table \ref{table:2}
and in the two $b$ asymmetries
of Figure \ref{fig:1}. This is mainly due to the new $q \bar q$ subprocess
appearing at the NLO.
In Table \ref{table:3} we assumed a $b$-tagging 
efficiency $\epsilon$ of $100\ \%$ and no contamination 
$\omega$ in disentangling $b$ and $\bar b$ jets. 
Taking $\epsilon$ and $\omega$ into account means in practice
dividing $\delta \sin ^2 \theta _{\rm eff}^{\rm lept}$ 
coming from the $Vb$ events by $\sqrt{\epsilon} (1-2\omega)$.
A typical realistic value for $\epsilon$ is $50\ \%$ while $\omega$ 
can be estimated as follows.
Once the forward and backward hemispheres are identified, event by event, with
some criterion, the charge separation
$\delta_b^{\exp}$ of the average charges measured in both hemispheres 
can be determined:
$ 
\delta_b^{exp}= <Q_b>_F -<Q_b>_B
$.  
A simple calculation yields, for small asymmetries, a relation among
$\delta_b^{exp}$, the bare quark charge $Q_b$ and $\omega$:
$  
\delta_b^{exp} = 2 Q_b (1-2\omega)
$. 
Using $Q_b=-\frac{1}{3}$, together with the experimental LEP value
$\delta_b^{\exp}=-0.21$ \cite{Abreu:1998nr}, gives $\omega \sim 0.34$.
This loss of precision is partly compensated by the fact that the Table
refers to $b$ production only; adding $\bar b$ doubles 
the available statistics.
At any rate approaching the quoted precisions will be
an experimental challenge.

Another source of uncertainty, that is not accounted for in
Table \ref{table:3}, is the dependence of the asymmetries
on the chosen set of parton densities. 
We investigated it by recomputing them with
different parton distribution sets in the classes 
{\tt cteq} and {\tt mrst}. 
By doing so, variations of the asymmetries of the order of 10\% 
can be easily observed around the $Z$ peak at both colliders.
This important dependence on the parton densities can be 
considered as an extra handle provided by the asymmetry measurements 
in constraining the parton distribution functions.
Conversely, with a more precise knowledge of them, the charge 
asymmetries can be used for precision measurements.

Although a complete NLO result is essential to predict the correct
production rates, a realistic experimental analysis is better 
performed with a tree level program allowing an 
easier interface with parton shower and hadronization packages \cite{ALPGEN}.
While this looks feasible in the $Vj$ case, the
stronger impact of the NLO corrections seems 
to prevent this possibility for the $Vb$ production process.
A possible way out is using a program as ALPGEN and produce
the $Vb$ final state only through the tree level 
$gg \to V b (\bar b)$ and $q \bar{q} \to V b (\bar b)$ subprocesses.
The former rate is finite when computed with $m_b \ne 0$.
As a result of such approximation one gets the dotted lines of
Figure \ref{fig:1}, that provide a good approximation 
to the exact asymmetries around the $Z$ peak.
As a drawback of this approach wrong rates are obtained, namely
$\sigma^{Vb}= 1.0$ ($0.04$) pb at LHC (Tevatron) to be compared with
the NLO numbers in Table \ref{table:3}. 
Therefore a $K$ factor should be included. 
A rather easy way to achieve this is using a fake $b$ mass.
We checked that with $m_b \sim 1$ GeV the correct $Vb$ production
rate is reproduced at Tevatron, leaving the corresponding
asymmetry nearly unchanged.


\begin{thebibliography}{99}
\bibitem{MCFM}   J.M. Campbell, R.K. Ellis, Phys. Rev. D {\bf 62}, 114012 (2000).
\bibitem{ALPGEN} M.L. Mangano {\em et al.}, JHEP {\bf 0307}, 001 (2003); see also R. Pittau, hep-ph/0209087.
\bibitem{delAguila:2005cn} F.~del Aguila, Ll.~Ametller and R.~Pittau, Phys.\ Lett.\ B {\bf 628}, 40 (2005).
\bibitem{Aguila02} F. del Aguila, Ll. Ametller and P. Talavera, Phys. Rev. Lett. {\bf 89}, 161802 (2002). 
\bibitem{Abreu:1998nr} P.~Abreu {\it et al.}, DELPHI Collaboration, Eur.\ Phys.\ J.\ C {\bf 9}, 367 (1999).
\end{thebibliography}
\end{document}